\documentstyle{article}
\begin{document}

\thispagestyle{empty}
\parskip=12pt
\raggedbottom

{\hfill  BUTP--97/02}\\
\vspace*{1cm}
\begin{center}
{\LARGE Cluster algorithms}
\footnote{Lecture given at the summer school on `Advances
in Computer Simulations', Budapest, July 1996. 
Work supported in part by Schweizerischer Nationalfonds}

\vspace{1cm}

Ferenc Niedermayer\footnote{On leave from the Institute of Theoretical
Physics, E\"{o}tv\"{o}s University, Budapest}
\\
Institut f\"{u}r theoretische Physik \\
Universit\"{a}t Bern \\
Sidlerstrasse 5, CH-3012 Bern, Switzerland

%\vspace{1.5cm}
%{December 1996} \vspace*{0.5cm}           % define a fixed point date

\nopagebreak[4]

\begin{abstract}
Cluster algorithms for classical and quantum spin systems are
discussed. In particular, the cluster algorithm is applied
to classical O(N) lattice actions containing interactions 
of more than two spins. The performance of the multi-cluster 
and single--cluster methods, and of the standard and improved 
estimators are compared.
\end{abstract}

\end{center}

\vspace{7cm}

\newpage            %\eject 

%======================================================================
\section{Introduction}
Simulations of statistical systems near to their critical points
is usually a very difficult problem, because the dynamical
evolution of the system in Monte Carlo (or computer-) time slows down 
considerably.
The phenomenon is called critical slowing down (CSD).
It also occurs in real systems (in real time) and is an interesting 
subject to study. 
In Monte Carlo (MC) simulations, however, where one is interested
in quantities averaged over the equilibrium probability distribution,
CSD could lead to a tremendous waste of computer time.
The evolution of a system is characterized by the autocorrelation time
--- this defines the rate how the system looses
memory of a previous state, or in other words, the number of MC steps
needed to generate a statistically independent new configuration.

Let us consider an observable $A$ (a function of spins if we consider
a spin system) and denote by $A_t$ its value at a given MC time $t$.
We define the {\em autocorrelation time} through the average over the
equilibrium distribution as
\begin{equation}
C_{AA}(t)= \langle A_s A_{s+t}\rangle -\langle A \rangle^2
\label{CAA}
\end{equation}
For large $t$ this decays exponentially, 
$C_{AA}(t) \propto \exp(-t/\tau_{exp,A})$ where $\tau_{exp,A}$, 
the {\em exponential autocorrelation time} corresponds to the slowest
mode in the MC dynamics. It is also useful to define a slightly
different quantity, the {\em integrated autocorrelation time}
\begin{equation}
\tau_{int,A}=\frac{1}{2}\sum_{t=-\infty}^{+\infty}C_{AA}(t)/C_{AA}(0)
=\frac{1}{2} + \sum_{t=1}^{+\infty}C_{AA}(t)/C_{AA}(0)
\label{tauint}
\end{equation}
This is the quantity which is relevant for the statistical error 
in a MC simulation. One usually estimates $\langle A \rangle$ from
the average of $n$ subsequent measurements, i.e. through the quantity
\begin{equation}
\bar{A}=\frac{1}{n}\sum_{t=1}^{n}A_t
\label{barA}
\end{equation}
Its statistical error, $\delta \bar{A}$ for $n\gg \tau_{int,A}$
is given by
\begin{equation}
\delta \bar{A}=\sqrt{2\tau_{int,A}} \cdot\delta\bar{A}_{naive}
\label{dA}
\end{equation}
where
\begin{equation}
\delta \bar{A}_{naive}=\sqrt{\frac{1}{n}C_{AA}(0)}
\label{dbarA}
\end{equation}
Equation (\ref{dA}) means that in order to reach a given accuracy 
one has to spend a computer time $\propto \tau_{int}$.
Unlike the equilibrium averages $\langle A \rangle$ or $C_{AA}(0)$,
the autocorrelation time depends on the actual MC algorithm.
In general, near a critical point (i.e. for $\xi \gg 1$) the
autocorrelation time diverges as
\begin{equation}
\tau \approx c \xi^z
\label{tau}
\end{equation}
where $z$ is the {\em dynamical critical exponent}.\footnote{For more 
details and for references on statistical time--series analysis see 
e.g. \cite{Madras}.}

For a local update, as the standard Metropolis algorithm, one has 
$z\approx 2$. Some 2d systems reveal their critical properties only at
quite large correlation length $\xi$. In two dimensions the memory
allows to consider systems with linear size $L\approx 1000$ and
a correlation length $\xi\approx 100 - 200$. For a local
algorithm, however, (\ref{tau}) predicts $\tau \sim 10000$,
which makes the simulation with a local Metropolis algorithm hopeless.

The $\tau\propto \xi^2$ behaviour reminds one of the the random walk 
--- a change at some site will propagate to a distance of $\sim \xi$ 
due to random local updating steps in a time proportional to $\xi^2$.

There are several ways to change the MC dynamics in order to decrease
the value of $z$. One way is to perform a deterministic local update
instead of a random one, staying on the energy surface $E$=const,
i.e. making a micro-canonical step. This {\em over-relaxation
algorithm} \cite{Adler} can lower the value of the dynamical
critical exponent down to $z\approx 1$.

The other class of MC dynamics updates a collective mode instead of
a single local variable. In some optimal cases one can reach 
$z\approx 0$, i.e. completely eliminate (at least from a practical 
point of view) the problem of CSD.
Two important methods of this type are
\begin{itemize}
\item Multi-grid algorithms
\item Cluster algorithms
\end{itemize}
In the multi-grid algorithms one introduces in addition to the original
(fine grid) lattice a sequence of lattices with lattice spacings
$a'=2a$, $a''=4a$, \ldots and updates the corresponding block of
spins. i.e. regions of given sizes and shapes. For the O($N$) vector
model the multi-grid algorithm reduces the dynamical critical exponent
to $z\approx 0.5 - 0.7$ \cite{Edwards0}.
As we shall see, the cluster algorithms -- the topics of
these notes -- work better for this system.
One should keep in mind, however, that for other systems the presently
available cluster algorithms are not efficient, while the multi-grid
algorithms can be still used.
In these lecture notes I discuss the cluster algorithms for classical
and quantum spin systems.
\section{Cluster Algorithm for Classical Spin Systems}

The cluster MC method has been suggested by Swendsen and Wang \cite{SW}
for the Ising model, based on an earlier observation that the partition
function of the system can be written as a sum over cluster
distributions \cite{FK}. In Ref.\,\cite{N1} the algorithm has been
generalized to spin systems, using more general considerations. Here I
follow this derivation since it will be straightforward to apply it 
to general O($N$) lattice actions.

Consider a general O($N$) vector model where the energy of the
configuration $S=\{ {\bf S}_n \}$ is given by\footnote{For simplicity,
we include the factor $\beta=1/kT$ in the definition of $E$, i.e. the
Boltzmann factor is $\exp(-E)$}
\begin{equation}
E(S)=\sum_l E_l(S)\enspace .
\label{ES}
\end{equation}
Here $l$ denotes a given link, a pair of sites $l=(n_1,n_2)$
and the interaction term $E_l(S)$ depends only on the corresponding 
spins, $E_l(S)=E_l({\bf S}_{n_1},{\bf S}_{n_2})$.
At this point let us extend the class of actions considered in
Ref.\,\cite{N1} and consider the case of 3-spin (4-spin, \ldots)
interaction terms. Let $l$ be a (hyper)link
$l=(n_1,n_2,n_3)$ and the corresponding interaction term
$E_l(S)=E_l({\bf S}_{n_1},{\bf S}_{n_2},{\bf S}_{n_3})$
(and similarly for 4-spin, \ldots interactions).
We shall also assume that the interaction has a global symmetry,
\begin{equation}
E_l(g{\bf S}_1,g{\bf S}_2,\ldots)=
E_l({\bf S}_1,{\bf S}_2,\ldots)\enspace ,
\label{gsep}
\end{equation}
for $g\in G$ where $G$ is the corresponding symmetry group.
To construct the clusters we connect the sites
belonging to some link $l$ by a `bond' with some probability 
$p_l(S)$ depending on the spins associated with the link.
We assume it to be globally invariant as well:
\begin{equation}
p_l(g{\bf S}_1,g{\bf S}_2,\ldots)=
p_l({\bf S}_1,{\bf S}_2,\ldots)\enspace .
\label{gsp}
\end{equation}
The probability to produce a given configuration of bonds $B$ is given
by
\begin{equation}
w(B|S)=\prod_{l\in B}p_l(S)\prod_{l\notin B}(1-p_l(S)) \enspace .
\label{wBS}
\end{equation}
The next step is to build the clusters. The cluster is the set of
sites which can be visited from each other going through bonds.
We propose now the following change in the configuration:
spins within a given cluster are transformed globally,
by some $g_i\in G$, $i=1,\ldots,N_{\rm C}$, where $N_{\rm C}$
is the actual number of clusters defined by the bond configuration.
(Note that different bond configurations can result in the same
cluster configuration. In fact, only the cluster configuration matters
here.)
The equation of detailed balance (assuming the same a priori
probability for $g_i^{-1}$ as for $g_i$) is given by
\begin{equation}
{\rm e}^{-E(S)} w(B|S) w(S\to  S' )=
{\rm e}^{-E( S' )} w(B| S' ) w( S' \to S)\enspace .
\label{dbgen}
\end{equation}
Here $w(S\to  S' )$ and $w( S' \to S)$ are the corresponding
acceptance probabilities used to correct the equation 
(see Ref.\,\cite{Krauth}).
Because of the global symmetry (\ref{gsep},\ref{gsp}),
the contributions coming from a link $l$ with all its sites 
belonging to the same cluster cancel and we obtain
\begin{equation}
\prod_{l\notin B} {\rm e}^{-E_l(S)} (1-p_l(S)) w(S\to S' )=
\prod_{l\notin B} {\rm e}^{-E_l( S' )} (1-p_l( S' ))
w( S' \to S)\enspace .
\label{db0}
\end{equation}
Let us introduce a convenient parametrization for $p_l(S)$:
\begin{equation}
p_l(S)=1-{\rm e}^{E_l(S)-Q_l(S)}\enspace .
\label{plS}
\end{equation}
The fact that $p_l(S)$ is a probability requires
\begin{equation}
Q_l(S) \ge E_l(S)\enspace .
\label{QlEl}
\end{equation}
Equation (\ref{db0}) gives then
\begin{equation}
\prod_{l\notin B} {\rm e}^{-Q_l(S)} w(S\to S' )=
\prod_{l\notin B} {\rm e}^{-Q_l( S' )} w( S' \to S)\enspace .
\label{db}
\end{equation}
Here in fact only those links contribute which would connect different
clusters, i.e. are on the `surface' of clusters.
We have succeeded to map our original spin system onto a system of
$N_{\rm C}$ `sites' (representing the clusters $1,\ldots,N_{\rm C}$)
with the dynamical variables $g_i, i=1,\ldots,N_{\rm C}$. The interaction
between the clusters is given by $Q_l(S)$ where $l$ is a (hyper)link
connecting two or more clusters. (Note that by taking $Q_l(S)=E_l(S)$,
i.e. $p_l(S)=0$, each lattice site becomes a cluster by itself and we
recover the local MC method.)

Let us specify the bond probabilities, i.e. $Q_l(S)$. 
We restrict the proposed transformations $g$ to a subgroup
$H\subset G$ to be specified later, and let $Q_l(S)$ be the maximal
energy $E_l( S' )$ where the elements $g\in H$ are applied independently
to all possible sites belonging to $l$,
\begin{equation}
Q_l(\vec{S}_1,\vec{S}_2,\ldots)=\max_{g_1,g_2,\ldots \in H}
E(g_1 \vec{S}_1, g_2 \vec{S}_2,\ldots)\enspace .
\label{Qdef}
\end{equation}
Clearly, we have $Q_l(g_1 \vec{S}_1, g_2 \vec{S}_2,\ldots)=
Q_l(\vec{S}_1,\vec{S}_2,\ldots)$
for any $g_1,g_2,\ldots\in H$.
For this choice of the bond probabilities the remaining factors
$l\notin B$ in (\ref{db}) cancel and one obtains
\begin{equation}
w(S\to S' )=w( S' \to S)\enspace .
\label{db1}
\end{equation}
This is a remarkable fact: the accept--reject step needed usually to
correct the equation of detailed balance is unnecessary, the clusters
can be updated independently. This is possible because the clusters 
are built dynamically, i.e. they are sensitive to the interaction
and to the actual configuration. With a fixed a priori given shape
of clusters one cannot achieve this.

The construction is still quite general, but there is a hidden subtle
point here. If $Q_l - E_l$ is too large then the bond probability
becomes too large as well. As a consequence, the largest cluster will
usually occupy almost the whole lattice, leaving out only a few small
isolated clusters. Although the algorithm is still correct,
it is also useless in this case: applying a global transformation to
the whole lattice does not change the relative orientation of the
spins, so one would update effectively a few small clusters with 
the computation being done on the whole lattice. 
Before proceeding with the general case of the O($N$) spin model, 
let us consider the Ising model as a
special case.
\begin{equation}
E=-J\sum_{n,\mu} S_n S_{n+\hat{\mu}}\enspace ,
\label{Ising}
\end{equation}
where $J>0$ and $S_n=\pm 1$.
The global symmetry transformation in this case is $S_n \to g S_n$,
$g=\pm 1$. Equation (\ref{Qdef}) gives
\begin{equation}
Q_l(S)=J\enspace ,
\label{QIsing}
\end{equation}
and consequently
\begin{equation}
p_l(S_1,S_2)= \left\{ \begin{array}{ll}
    1-{\rm e}^{-2J} & \mbox{ for parallel spins,  $S_1 S_2=+1$}\enspace , \\
    0               & \mbox{ for antiparallel spins, $S_1 S_2=-1$}\enspace .
    \end{array}
\right.   
\label{pIsing}
\end{equation}

This is the bond probability for the Swendsen--Wang algorithm.
An important observation is that the clusters cannot grow too large:
since the bond probability is zero between antiparallel spins,
all the spins in a cluster will have the same sign -- the clusters
cannot grow larger than the region of same-sign spins. In the unbroken
phase the latter does not exceed half of the total number of sites,
hence the main danger of the cluster algorithms -- that one ends up
with a single cluster occupying nearly the whole volume -- is avoided
in this case.

Since the clusters do not interact (c.f. (\ref{db1})) one can choose
any of the $2^{N_{\rm C}}$ sign assignments with equal probability.
One can average over these $2^{N_{\rm C}}$ possibilities without actually 
doing the updates -- by introducing the {\em improved estimators}.
For the correlation function one has
\begin{equation}
\langle S_x S_y \rangle = \sum_{\cal C} p({\cal C}) 
\langle S_x S_y \rangle_{\cal C} \enspace ,
\label{SxSy}
\end{equation}
where ${\cal C}$ is a given cluster distribution appearing with
probability $p({\cal C})$ and $\langle S_x S_y \rangle_{\cal C}$
is the average over the $2^{N_{\rm C}}$ possibilities for ${\cal C}$.
The latter is trivial,
\begin{equation}
\langle S_x S_y \rangle_{\cal C} = \delta_{xy}({\cal C})\enspace ,
\end{equation}
where $\delta_{xy}({\cal C})$ is $1$ if the two sites belong to the
same cluster, and $0$ otherwise.
Hence we have
\begin{equation}
\langle S_x S_y \rangle = \langle \delta_{xy}({\cal C}) \rangle\enspace .
\end{equation}
The improvement comes from the fact that at large distances $|x-y|$
the small value of correlator $\langle S_x S_y \rangle \ll 1$
is obtained by averaging over values $+1$ and $-1$ in the standard
measurement while in the case of the improved estimator
$\delta_{xy}({\cal C})$ it comes from averaging over $+1$ and $0$.
In other words, the value of $C_{AA}(0)$ in (\ref{dbarA}) is much
smaller for the improved estimator than for the original one since
\begin{equation}
\langle (S_x S_y)^2 \rangle =1, ~~~
\langle (\delta_{xy}({\cal C}))^2 \rangle =
\langle \delta_{xy}({\cal C}) \rangle \ll 1 \enspace .
\label{var}
\end{equation}
As a consequence, one expects to gain an additional factor
$1/\langle S_x S_y \rangle$ in computer time. 
This argument is, unfortunately, not complete. The point is that
one usually measures the correlation function $\langle S_x S_y \rangle$
from an average over the whole lattice with $|x-y|$ fixed.
When the site $x$ runs over the lattice the quantity $S_x S_y$
fluctuates more independently than the improved estimator
$\delta_{xy}({\cal C})$ -- the values of the latter are strongly
correlated when large clusters are present.

A variant of this algorithm has been introduced by Wolff \cite{Wolff1}
-- the single-cluster algorithm. He suggested to build just one
cluster starting from a random site, reflect all spins in this cluster
and start the procedure again. The difference between this and the
original Swendsen--Wang multi-cluster algorithm is that one enhances
the probability of updating large clusters since a cluster of size
$|C|$ is hit with the probability $|C|/V$, where $V$ is
the total number of lattice sites. Large clusters evolve still too
slowly compared to small ones in the multi-cluster method --
the single-cluster version corrects for this by updating more often
the large clusters. (Obviously, one can try to vary the maximal
number of clusters to be updated within a given MC run, and optimize
in the distribution of this number.)

The improved estimator can be modified for the case of the
single-cluster algorithm \cite{Wolff1a}. For example, the correlation
function is given by
\begin{equation}
\frac{1}{V}\sum_{x=1}^V \langle S_x S_{x+r} \rangle =
\left\langle \sum_{i=1}^{N_{\rm C}} \frac{|C_i|}{V}\,\frac{1}{|C_i|}
\sum_{x\in C_i} \delta_{x,x+r}({\cal C})\right\rangle^{\rm \!\!mc}
\!\!= \left\langle \frac{1}{|C_1|}\sum_{x\in C_1}
\delta_{x,x+r}({\cal C})\right\rangle^{\!\!\rm sc},
\end{equation}
where the superscripts mc, sc refer to multi-cluster and single-cluster
algorithms, respectively.
The susceptibility in the unbroken phase is given by
\begin{equation}
\chi=\frac{1}{V}\sum_{x,y}\langle S_x S_y \rangle =
\left\langle \frac{1}{V}\sum_{i=1}^{N_{\rm C}} \sum_{x,y\in C_i} 
\delta_{xy}({\cal C}) \right\rangle \!\!=
\left\langle \sum_{i=1}^{N_{\rm C}} \frac{|C_i|}{V} |C_i| \right\rangle =
\left\langle |C| \right\rangle^{\!\!\rm sc}  .
\end{equation}
The average size of a randomly chosen cluster diverges together with
the susceptibility when one approaches the critical point -- the MC
dynamics adjusts itself to the long range correlations present
in the equilibrium situation.

In applying the previous considerations to the O($N$) spin model
one has to choose the proper symmetry transformation $H$ in
(\ref{Qdef}). Choosing this to be the rotation group 
-- the choice made in \cite{N1} -- leads to a bond
probability which is too large, the largest cluster tends to occupy
the whole lattice. It has been suggested there to make only
small rotations and  consider the case of interacting clusters
as in (\ref{db}).
A much better solution suggested by Wolff \cite{Wolff1}
is to choose the subgroup of reflections with respect to 
some given random direction $\vec{r}$. 
(The derivation in Ref.~\cite{Wolff1} was based on embedding an 
Ising model into the O($N$) model and applying to it the Swendsen-Wang
cluster algorithm. The advantage of the present approach shows up
in applications to O($N$) lattice actions of more general type.)
The corresponding subgroup contains the identity and the reflection
of the parallel component of the spin to the vector $\vec{r}$, i.e.
\begin{equation}
g:~~ S_n^{\parallel} \to -S_n^{\parallel}\enspace ,
~~~ \vec{S}_n^{\perp} \to \vec{S}_n^{\perp} \enspace .
\end{equation}
For the standard O($N$) action we have
\begin{equation}
\vec{S}_n \vec{S}_{n+\hat{\mu}} =
S_n^{\parallel} S_{n+\hat{\mu}}^{\parallel} +
\vec{S}_n^{\perp} \vec{S}_{n+\hat{\mu}}^{\perp}\enspace ,
\end{equation}
and only the first term is affected by the update.
Essentially we obtained an embedded Ising model with space-dependent
{\em ferromagnetic} couplings,
\begin{equation}
E=-\sum_{n,\mu} J_{n,\mu} \epsilon_n \epsilon_{n+\hat{\mu}}\enspace ,
\mbox{~~~ with ~} J_{n,\mu}= 
J |S_n^{\parallel} S_{n+\hat{\mu}}^{\parallel}|, ~~
\epsilon_n = {\rm sign} S_n^{\parallel}\enspace .
\end{equation}
The corresponding bond probabilities are given by $Q_l=J_l$.
Again, because of the ferromagnetic nature of the couplings, 
the regions with $\epsilon_n > 0$ and $\epsilon_n <0$ are not
connected by bonds and the size of the clusters is bounded by the size
of the corresponding regions. Consequently (at least in the unbroken
phase)  one is safe from having clusters with $|C|\sim V$.
For the O($N$) vector model both the single-cluster \cite{Wolff3}
and multi-cluster \cite{Edwards} method eliminates practically the CSD.
For this model the cluster algorithm works even better than for 
the Ising model where some CSD is still observed.

Note that the present formulation can be easily applied to more
general O($N$) actions. Let us consider first the Symanzik 
(tree-level) improved O($N$) action \cite{Symanzik},
\begin{equation}
E(S)=-J\sum_{n,\mu} \left[
 \frac{4}{3} \vec{S}_n \vec{S}_{n+\hat{\mu}}
- \frac{1}{12} \vec{S}_n \vec{S}_{n+2\hat{\mu}} \right]\enspace .
\label{Sym}
\end{equation}
Here we have two types of links, $l=(n,n+\hat{\mu})$ and
$l=(n,n+2\hat{\mu})$, with 
$Q_l=\frac{4}{3} |S_n^{\parallel} S_{n+\hat{\mu}}^{\parallel}|$
and
$Q_l=\frac{1}{12} |S_n^{\parallel} S_{n+2\hat{\mu}}^{\parallel}|$,
respectively.
A potential danger is related to the fact that the second term
in (\ref{Sym}) is anti-ferromagnetic -- the bonds of the second type
can connect sites with opposite sign of $S_n^{\parallel}$.
Fortunately, the coefficient $1/12$ is sufficiently small, and
the largest cluster does not grow too large \cite{HN2}.

Another example of practical importance is provided by the classically
perfect lattice action (or the fixed point action of a renormalization
group transformation) \cite{HN:FP} which has a form
\begin{equation}
 E(S)=\frac{1}{2}\sum_{n,r}\rho(r) (1-\vec{S}_n \vec{S}_{n+r}) + 
\sum_{n_1..n_4} c(n_1..n_4)
(1-\vec{S}_{n_1} \vec{S}_{n_2})(1-\vec{S}_{n_3} \vec{S}_{n_4}) +\ldots
\label{FPaction}
\end{equation}
This lattice action is used to minimize the lattice artifacts (the
discretization errors). Although introducing the signs
$\epsilon_n={\rm sign}(S_n^{\parallel})$ does not turn this action
into an Ising model (couplings between more than two spins are also
present) the discussion presented here readily applies to this 
general action, and the procedure eliminates practically the CSD.

One can also introduce an external magnetic field. The simplest way to
apply the cluster algorithm is to consider the external field as an
extra spin which is coupled to all spins, and consider this
interaction term on the same footing as all other terms in 
$E(S)$\cite{Dimitrovic}.
The cluster to which the external spin belongs does not need to be
updated, otherwise the procedure is unchanged.

When modifying the algorithm, the condition of detailed balance
has to be rechecked carefully. To illustrate this, let us consider
the 3d O(3) model in the broken phase.
If the random direction $\vec{r}$ points towards the magnetization 
$\vec{M}$, the effective couplings 
$J|S_n^{\parallel}S_{n+\hat{\mu}}^{\parallel}|$ 
become too large, consequently also the size of the largest cluster.
(This is expected physically -- flipping half of the spins along
the total magnetization does not produce a typical configuration.)
Could one modify the algorithm by restricting $\vec{r}$ to be
orthogonal to $\vec{M}$? It is easy to see that this is not correct. 
The component $\vec{M}^{\perp}$ (orthogonal to $\vec{r}$)
remains unchanged by the updates while $M^{\parallel}$ changes from
zero to some nonzero value hence $|\vec{M}|$ always grows.
In fact, we violated detailed balance by biasing $\vec{r}$ with the
direction of the magnetization. (On the other hand, it is allowed to
take $\vec{r}$ orthogonal to the direction of a given spin.)

One can also define improved estimators for the O($N$) spin model.
Since only the signs of $S_n^{\parallel}$ are updated one can use
the relation \cite{Wolff3}
\begin{equation}
\langle \vec{S}_x \vec{S}_y \rangle =
N \langle S_x^{\parallel} S_y^{\parallel} \rangle \enspace ,
\end{equation}
and define the improved estimator as
\begin{equation}
N |S_x^{\parallel} S_y^{\parallel}| \, \delta_{xy}({\cal C})\enspace .
\end{equation}
Unfortunately, this procedure introduces an unwanted noise in the
correlator when $\vec{S}_x$ and $\vec{S}_y$ are correlated
significantly (i.e. for $|x-y| \le \xi$). To illustrate this let us
consider  $\vec{S}_x \vec{S}_x$: this is, of course, exactly 1 when
measured directly, it has no statistical error while the estimator 
$N S_x^{\parallel} S_x^{\parallel}$ fluctuates.
The problem is cured easily \cite{N2}: one should only choose $N$
random directions $\vec{r}_1,\vec{r}_2,\ldots,\vec{r}_N$
in a sequence forming an {\em orthogonal system} and apply the cluster
algorithm using $\vec{r}=\vec{r}_i$ for $i=1,\ldots,N$, otherwise 
everything remains unchanged, including the improved estimator.
Unfortunately, for the single-cluster update this trick is not very
convenient. To cure the problem with the improved measurement and keep
the advantage that the single-cluster method updates more efficiently 
the large clusters one can do the following:
\begin{enumerate}
\item make $n_{\rm s}$ single-cluster steps, but do not measure anything,
\item make $N$ consecutive multi-cluster steps with a random basis
and perform a measurement after each step.
\end{enumerate}
One can try to optimize the value of $n_{\rm s}$ to get a uniformly
small autocorrelation time at all scales or even try to vary the
maximal number of clusters ($1,2,\ldots$) to be built and updated.
We shall refer to this as the `hybrid' method. 
To illustrate the performance of the cluster method and of the
improved estimators I made some runs\footnote{Most of these runs were 
made after the school, while preparing this written version.}
on a $256^2$ lattice at $\beta=1.8$ which corresponds to a correlation
length $\xi=64.78(15)$ \cite{Wolff3}. For the hybrid algorithm I took
$n_{\rm s}=25 \approx V/ \langle |C| \rangle$, and measured the
correlation function $C(x)=\langle \vec{S}_0 \vec{S}_x \rangle$
using the standard estimator (averaged over the lattice volume)
and the improved estimator. The measurements in the single-cluster
method were done after each $n_{\rm s}$ updates.
Table~1 shows the results for $C(x)$ at $x=1,20,60,120$, and for the
susceptibility. To compare the performance the squared errors are
shown, normalized to 20000 sweeps for each case.

\begin{table}[htb]
%\begin{flushleft}
\begin{center}
\renewcommand{\arraystretch}{1.2}
\begin{tabular}{ccccccc}
\hline\noalign{\smallskip}
method & $C(1)$ & $C(20)$ & $C(60)$ & $C(120)$ & $\chi$ \\
\noalign{\smallskip}\hline\noalign{\smallskip}
H, imp. & 0.68796(3) & 0.1973(2) & 0.0731(3) & 0.0354(4) 
                                                        & 3560(20) \\
\noalign{\smallskip}\hline\noalign{\smallskip}
S, st. & $2\cdot 10^{-9}$ & $5\cdot 10^{-8}$ & $1\cdot 10^{-7}$ &
 $2\cdot 10^{-7}$ &  $6\cdot 10^{2}$ & \\
M, st. &  $1.6\cdot 10^{-9}$ & $3\cdot 10^{-7}$ & $9\cdot 10^{-7}$ &
 $1\cdot 10^{-6}$ &  $2\cdot 10^{3}$ \\
H, st. & $1.3\cdot 10^{-9}$ & $9\cdot 10^{-8}$ & $3\cdot 10^{-7}$ &
 $4\cdot 10^{-7}$ & $7\cdot 10^{2}$ \\
\noalign{\smallskip}\hline\noalign{\smallskip}
S, imp. &  $2\cdot 10^{-5}$ & $5\cdot 10^{-6}$ & $2\cdot 10^{-6}$ &
 $7\cdot 10^{-7}$ &  $3\cdot 10^{3}$ \\
M, imp. &  $1\cdot 10^{-9}$ & $1\cdot 10^{-7}$ & $3\cdot 10^{-7}$ &
 $4\cdot 10^{-7}$ & $1\cdot 10^{3}$ \\
H, imp. &  $1\cdot 10^{-9}$ & $5\cdot 10^{-8}$ & $1\cdot 10^{-7}$ &
 $1.4\cdot 10^{-7}$ & $4\cdot 10^{2}$ \\
\noalign{\smallskip}\hline\noalign{\smallskip}
H, imp.${}^*$ &  $7\cdot 10^{-10}$ & $2\cdot 10^{-8}$ &
 $5\cdot 10^{-8}$ &  $6\cdot 10^{-8}$ & $1.5\cdot 10^{2}$ \\
\noalign{\smallskip}\hline
\end{tabular}
\renewcommand{\arraystretch}{1}
%\end{flushleft}
\end{center}
\caption[]{Spin-spin correlation function and susceptibility
with different variants of the cluster algorithm (single-cluster,
multi-cluster and `hybrid'), measured with
standard and improved estimators. The first line shows the data,
the following lines the error squared, normalized to 20\,ksweeps.
The last line shows the (impractical) choice when the multi-cluster
steps are used only in the improved estimator but not in updating
the configuration.}
\end{table}

Comparing the standard estimators one sees that the
single-cluster method updates all scales better, except for the
shortest distances. The error squared for the hybrid method is 
somewhat larger because we made there 
$\sim 1$ single-cluster sweep (i.e. $n_{\rm s}=25$ single-cluster
updates) followed by 3 multi-cluster updates with measurements.
A striking observation is that for the single-cluster method 
the standard estimator produces
a smaller error than the improved estimator -- this is naturally
connected with the noise discussed above. For the multi-cluster or
hybrid method the corresponding improved estimator (which
cancels this noise by taking a random orthogonal basis) has 
obviously smaller errors than the standard estimator, but the
gain is about a factor of $2-3$, much smaller than predicted by 
the naive argument using the relations in (\ref{var}). 
At $x=60\approx \xi$ the hybrid version gives an improvement 
in computer time of a factor $3$ compared to the multi-cluster 
version and a factor of $20$ compared to the single-cluster version
with (noisy) improved estimator. It seems that the best strategy is 
to use the single-cluster method with the {\em standard estimator}
or the hybrid version with the improved estimator.
Of course, at this correlation length all these versions of the
cluster method are incomparably better than the local Metropolis update. 

To separate the effect of multi-cluster updates from 
that of the multi-cluster improved estimator I made a run where after 
$n_{\rm s}$ single cluster updates in 3 consecutive steps 
(in all 3 orthogonal directions) the multi-clusters were constructed
and the corresponding improved estimators were measured 
but the resulting multi-cluster configurations were not updated. 
Accordingly, only the single-cluster steps were counted in the total 
number of sweeps. These results are shown in the last line, as 
H,imp.${}^*$.
Note that this is not a practical choice since it takes no time 
to update the clusters once they were constructed.
Although we did not count the construction of multi-clusters 
(since it does not affects the configurations) it counts,
of course, in computer time.

The cluster method has still many interesting applications for the
O($N$) vector model [18--23]. As an example, let me mention
here the determination of the interface tension in the Ising model
by Hasenbusch \cite{Hasenbusch1}. He considered the Ising model
on a 2d strip with different (periodic and anti-periodic) boundary
conditions in the time direction. The boundary condition has been
considered as a dynamical variable\footnote{A similar trick has been
used in \cite{HHN} for SU(3) gauge theory to determine the string
tension.}, $\epsilon=1$ for periodic and $\epsilon=-1$ for 
anti-periodic b.c. The idea in Ref.~\cite{Hasenbusch1}
was that when a line of deleted bonds\footnote{Sometimes it is more
convenient to say that one puts originally bonds everywhere and then
deletes them with probability $\bar{p}_l=1-p_l$.} cuts through the
strip in spatial direction one can flip (with probability 1) 
the region between this line and the boundary (at $t=0$) flipping 
the sign of $\epsilon$ at the same time. 
The relative number of cases with $\epsilon=1$ and $-1$
is related to the free energy of a kink, i.e. to the interface
tension.

Unfortunately, the cluster algorithm does not work well for other
models \cite{Sokal2}, at least in the most convenient version with
independent clusters. The reason is always the same: the largest
cluster tends to occupy the whole volume. To illustrate this,
consider the Ising spin-glass,
\begin{equation}
E(S)=-\sum_{\langle i,j \rangle} J_{ij} S_i S_j\enspace ,
\end{equation}
where the random couplings can be positive or negative. In this case
(when there are frustrated links) the clusters can grow through the
boundaries separating the regions with $S=1$ and $S=-1$, and for the
couplings of interest one big cluster is formed.
One can still give up the requirement of independent clusters and go
back to (\ref{db}). By decreasing $Q_l(S)$ one can make the clusters 
smaller -- the price paid for this is that they start to interact 
and the acceptance probability may be extremely small for clusters
of reasonable size. This possibility, however, has not been properly
investigated yet.

\section{Quantum Spin Systems}

We shall consider the example of the 2d anti-ferromagnetic quantum
Heisenberg model given by the Hamiltonian
\begin{equation}
H=J\sum_{x,\mu} \vec{S}_x \vec{S}_{x+\hat{\mu}}\enspace ,
\label{AFH}
\end{equation}
where $J>0$ and $\vec{S}_x=\frac{1}{2}\vec{\sigma}_x$ is 
a spin operator ($\vec{\sigma}_x$ are Pauli matrices) at site $x$. 
This old model seems to describe the dynamics of the electron spins 
within the copper-oxygen planes of the ${\rm La}_2{\rm CuO}_4$ material.
(Note that the first high-${\rm T}_c$ super-conductor discovered was
${\rm La}_{2-x}{\rm Ba}_x{\rm CuO}_4$ with the doping 
$x\approx 0.15$ \cite{hightc}.)

To simulate a quantum spin system is much more difficult than 
a classical one because -- as we shall see -- the MC updates
have to satisfy some constraints. Based on the work of Evertz, Lana and
Marcu \cite{Evertz} who developed a loop cluster algorithm for vertex
models, Wiese and Ying \cite {Wiese1} worked out an analogous 
procedure for the Hamiltonian (\ref{AFH}). I shall follow their 
derivation, with a few insignificant simplifications. 
In particular, I will consider the 1d case, and the multi-cluster 
version instead of the single-cluster one.

First we decompose the Hamiltonian into non-interacting sub-lattices,
$H=H_1+H_2$, where
\begin{equation}
H_1=J\sum_{n}\vec{S}_{2n} \vec{S}_{2n+1}\enspace ,~~~
H_2=J\sum_{n}\vec{S}_{2n-1} \vec{S}_{2n}\enspace ,
\label{H12}
\end{equation}
and use the Suzuki--Trotter formula for the partition function:
\begin{equation}
Z={\rm Tr}{\rm e}^{-\beta H}=
\lim_{N\to\infty}{\rm Tr}\left[{\rm e}^{-\epsilon\beta H_1}
{\rm e}^{-\epsilon\beta H_2} \right]^N= 
\lim_{N\to\infty}{\rm Tr}\left(
{\rm e}^{-\epsilon\beta H_1}{\rm e}^{-\epsilon\beta H_2}
%{\rm e}^{-\epsilon\beta H_1} 
\cdots \right)\enspace ,
\label{ST}
\end{equation}
where $\epsilon=1/N$ determines the lattice spacing in the Euclidean
time direction. By inserting a complete set of eigenstates
$|+\rangle$ and $|-\rangle$ of $\sigma_x^3$ between the factors
$\exp(-\epsilon\beta H_i)$ we obtain a classical system with Ising
like variables $S(x,t)=\pm 1$ at each site of the 2d lattice.
To see the structure of the corresponding contribution it is
sufficient to consider a single interaction term in (\ref{H12}),
$h=J\left(\vec{S}_a \vec{S}_b -\frac{1}{4}\right)$ where the term $1/4$ is
subtracted for convenience.
Since $\vec{S}_a \vec{S}_b -\frac{1}{4}=
\frac{1}{2}(\vec{S}_a + \vec{S}_b)^2-1$, the eigenvalues of
$h$ are $-J$ for the singlet state and $0$ for the triplet
state of the total angular momentum $\vec{S}_a + \vec{S}_b$. 
This gives the following transition amplitudes\footnote{A direct 
calculation leads to a negative number, $-w_3$ for the third
amplitude. Redefining the sign of the eigenvector $|-\rangle$
at every odd site makes this amplitude positive without
affecting the others.} for the operator
$A=\exp(-\epsilon\beta h)$:
\begin{equation}
\begin{array}{l}
 \langle ++|A|++\rangle =\langle --|A|--\rangle \equiv w_1=1\enspace , \\
 \langle +-|A|+-\rangle =\langle -+|A|-+\rangle \equiv w_2=
     \frac{1}{2}\left({\rm e}^{\epsilon\beta J}+1\right)   \enspace , \\
 \langle +-|A|-+\rangle =\langle -+|A|+-\rangle \equiv w_3=
     \frac{1}{2}\left({\rm e}^{\epsilon\beta J}-1\right)   \enspace . 
\end{array}
\label{amp0}
\end{equation}
All other matrix elements are zero:
\begin{equation}
\langle ++|A|+-\rangle \equiv w_4=0\enspace ,\ldots \\
\end{equation}
The separation $H=H_1+H_2$ leads to a checker-board type
interaction
of the corresponding classical spin system -- only the four spin
on the corners of `black' squares interact, producing the
corresponding factor $w_i$ in the Boltzmann factor.
The partition function is given by
\begin{equation}
Z=\sum_{\{S\}} w_1^{n_1(S)} w_2^{n_2(S)} w_3^{n_3(S)} w_4^{n_4(S)}\ldots
\label{Zw}
\end{equation}
where $n_i(S)$ is the number of `black' plaquettes of type $i$.
Clearly, only those configurations contribute for which
$n_4(S)=n_5(S)=\ldots =0$, i.e. the configurations should satisfy
the corresponding constraints. This requirement causes a serious
problem for a local updating procedure since most steps are forbidden.
In the loop cluster approach this problem is avoided -- one builds
closed loops of bonds on the configuration, and flips all spins
along a loop. The new configuration automatically satisfies the
constraints, and the bond probabilities are chosen to satisfy detailed
balance.

The first step is to connect the sites of a `black' square by bonds
with some probabilities depending on the type of the configuration.
The prescription is given in Table~2 (where the time axis is
the vertical one):

% definition of pictures:
\def\pav{
\begin{picture}(16,10)(0,8)
\linethickness{1pt}
\put(0,2){{\bf $+$}}
\put(15,2){{\bf $+$}}
\put(0,12){{\bf $+$}}
\put(15,12){{\bf $+$}}
\put(4,3){\line(0,1){10}}
\put(14,3){\line(0,1){10}}
\put(4,3){\circle*{1}}
\put(14,3){\circle*{1}}
\put(4,13){\circle*{1}}
\put(14,13){\circle*{1}}
\end{picture}
}
\def\pbv{
\begin{picture}(16,10)(0,8)
\linethickness{1pt}
\put(0,2){{\bf $+$}}
\put(15,2){{\bf $-$}}
\put(0,12){{\bf $+$}}
\put(15,12){{\bf $-$}}
\put(4,3){\line(0,1){10}}
\put(14,3){\line(0,1){10}}
\put(4,3){\circle*{1}}
\put(14,3){\circle*{1}}
\put(4,13){\circle*{1}}
\put(14,13){\circle*{1}}
\end{picture}
}
\def\pbh{
\begin{picture}(16,16)(0,8)
\linethickness{1pt}
\put(0,2){{\bf $+$}}
\put(15,2){{\bf $-$}}
\put(0,12){{\bf $+$}}
\put(15,12){{\bf $-$}}
\put(4,3){\line(1,0){10}}
\put(4,13){\line(1,0){10}}
\put(4,3){\circle*{1}}
\put(14,3){\circle*{1}}
\put(4,13){\circle*{1}}
\put(14,13){\circle*{1}}
\end{picture}
}
\def\pch{
\begin{picture}(16,16)(0,8)
\linethickness{1pt}
\put(0,2){{\bf $+$}}
\put(15,2){{\bf $-$}}
\put(0,12){{\bf $-$}}
\put(15,12){{\bf $+$}}
\put(4,3){\line(1,0){10}}
\put(4,13){\line(1,0){10}}
\put(4,3){\circle*{1}}
\put(14,3){\circle*{1}}
\put(4,13){\circle*{1}}
\put(14,13){\circle*{1}}
\end{picture}
}

\setlength{\unitlength}{1mm}
\begin{table}
\renewcommand{\arraystretch}{1.2}
\begin{center}
\begin{tabular}{ccl|ccl}
%\hline\noalign{\smallskip}
\hline
\rule[-0.5ex]{0pt}{3ex}
type & bonds & prob. & type & bonds & ~~~~prob. \\[1mm]
\hline
1 & ~~\pav ~~  & ~~~1 &
2 & ~~\pbv ~~  & ~~~$p=1/w_2$ \\
3 & ~~\pch ~~  & ~~~1 & 
2 & ~~\pbh ~~  & ~~~$p'=w_3/w_2$ \\
~  & & & & \\
~  & & & & \\
\hline
\end{tabular}
\renewcommand{\arraystretch}{1}
\caption[]{Bond probabilities for different spin 
configurations on a plaquette. Note that 
$p+p'=1$ for configurations of type 2.}
\end{center}
\end{table}

It is easy to see that on a finite periodic lattice these bonds form a
set of closed loops. Flipping all spins along a closed loop leads to a
new admissible configuration (i.e. no forbidden `black' plaquettes
of type $i=4,5,\ldots$ appear). By flipping the spins on one of the bonds 
of a plaquette, the transition probabilities are
$p(1\to 2)=1$, $p(2\to 1)=1/w_2=w_1/w_2$, $p(2\to 3)=w_3/w_2$,
$p(3\to 2)=1$.
These probabilities satisfy the relations
\begin{equation}
w_1 p(1\to 2)=w_2 p(2\to 1)\enspace , ~~~ 
w_2 p(2\to 3)=w_3 p(3\to 2)\enspace ,
\end{equation}
and consequently the condition of detailed balance for the equilibrium
distribution corresponding to (\ref{Zw}).
This technique made it possible to determine the low energy effective
parameters (as the spin wave velocity and spin stiffness) of 
the anti-ferromagnetic quantum Heisenberg model to a high precision 
\cite{Wiese1}. 

In (\ref{ST}) one has to take the $\epsilon\to 0$ ($N\to\infty$)
continuum limit in the time direction.
Notice that the only non-deterministic choice is for plaquette 
of type 2 where the possibility to put the link horizontally
(along the spatial direction) is 
\begin{equation}
p'=\frac{w_3}{w_2}=
\frac{{\rm e}^{\epsilon\beta J}-1}{{\rm e}^{\epsilon\beta J}+1}=
\frac{1}{2}\beta J \epsilon +{\rm O}(\epsilon^2) \enspace .
\end{equation}
One can take now a continuum time and reformulate the prescription 
into a continuum language: to turn a path from vertical into
horizontal direction in time interval $dt$ equals to 
$\frac{1}{2}\beta J dt$. 
Modifying the loop cluster algorithm this way 
Beard and Wiese \cite{Beard} have shown that this observation allows 
to design a loop cluster algorithm free from discretization error.

There are still many promising application of the cluster algorithm
to quantum systems but it is beyond our scope to discuss them here.
I would like to mention only a recent suggestion by Galli
\cite{Galli} to overcome the `sign problem'.

\section{Conclusion}

Cluster algorithms work excellently for classical and quantum spin
systems.  The way of choosing the collective modes to be updated
depends on the action and the actual configuration, and adjusts itself
optimally. The Boltzmann weight (in the optimal case) is completely
absorbed in the bond probabilities, hence the clusters can be updated
independently. This leads to an effective update of large scale
collective modes, to a strong reduction or even elimination of the 
critical slowing down, and to a possibility to introduce improved
estimators with reduced variance.
%


\end{document}